\documentclass{article}
\usepackage{spconf,amsmath,graphicx}

\usepackage{xcolor}
\usepackage{amsfonts}
\usepackage{bm}
\newcommand{\eqdef}{\stackrel{\textrm{\tiny def}}{=}}
\newcommand{\e}{\mathrm{e}}
\newcommand{\BM}{\text{\tiny B,M}}
\newcommand{\BR}{\text{\tiny B,R}}
\newcommand{\RM}{\text{\tiny R,M}}
\newcommand{\R}{\text{\tiny R}}
\newcommand{\BS}{\text{\tiny BS}}
\newcommand{\RIS}{\text{\tiny RIS}}
\newcommand{\ML}{\text{\tiny ML}}
\newcommand{\RML}{\text{\tiny RML}}

\usepackage{cite}

\newcommand\Mark[1]{\textsuperscript{#1}}

\title{RIS-aided Joint Localization and Synchronization with a Single-Antenna MmWave Receiver}
\name{Alessio Fascista\Mark{*}, Angelo Coluccia\Mark{*}, Henk Wymeersch\Mark{$\dagger$}, Gonzalo Seco-Granados\Mark{$\mathsection$}}
\address{\Mark{*}Department of Innovation Engineering, University of Salento, Italy\\ \Mark{$\dagger$}Department of Electrical Engineering, Chalmers University of Technology, Sweden\\ \Mark{$\mathsection$}Department of Telecommunications and Systems Engineering, Universitat Autonoma de Barcelona, Spain}
\begin{document}
\ninept
\maketitle
\begin{abstract}
MmWave multiple-input single-output (MISO) systems using a single-antenna receiver are regarded as promising solution for the near future, before the full-fledged 5G MIMO will be widespread. 
However, for MISO systems synchronization cannot be performed jointly with user localization unless two-way transmissions are used.
In this paper we show that thanks to the use of a reconfigurable intelligent surface (RIS), joint localization and synchronization is possible with only downlink MISO transmissions. The direct maximum likelihood (ML) estimator for the position and clock offset is derived.
To obtain a good initialization for the ML optimization,  a decoupled, relaxed estimator of position and delays is also devised, which does not require knowledge of the clock offset. 
Results show that the proposed approach attains the Cram\'er-Rao lower bound even for moderate values of the system parameters.

\end{abstract}
\section{Introduction}
\label{sec:intro}

The role of positioning in mobile communications has enormously grown in the fifth generation (5G) of cellular systems, due, among other factors, to to the use of mmWave massive multiple-input multiple-output (MIMO) technologies. 
However, before full-fledged MIMO infrastructures will be widespread, mmWave MISO systems using a single-antenna mobile station (MS) are regarded as a promising solution to obtain some of the benefits envisioned for  5G  already in the very near future \cite{Westberg}.
In fact, although the angle-of-arrival (AOA) information is not available in the MISO setup,  precise estimation of location-related information such as time-of-flight (TOF) and angle-of-departure (AOD) is still possible 
\cite{FascioMultiBS,MultiBS_nosynch_onlyAOD},  enabling accurate localization \cite{Lohan, Peral,Kakkavas}.

More specifically, mmWave MISO systems have been shown to achieve centimeter-level accuracy even for a single base station (BS) \cite{Fascio1}. Improved performance can be obtained when the transmit beamforming at the BS is adaptively steered towards the MS \cite{Fascio2}. 
Further studies demonstrate that, in addition to precise localization, the information brought by non-line-of-sight (NLOS) paths can be fruitfully leveraged in mmWave MISO systems to estimate the position of the physical scatterers or reflectors, thus enabling simultaneous localization and mapping of the radio propagation environment over time \cite{Fascio3}.

The use of TOF information  requires that the BS and the MS are synchronized. However, in case of a single BS --- which is seen as a common scenario for 5G picocells  --- the synchronization task cannot be performed jointly with the user localization using one-way transmissions. In particular, the ambiguity introduced by the clock offset cannot be resolved, which implies that the joint Fisher information matrix (FIM) in this setup becomes singular. The standard way to cope with this issue is adopt a  conventional two-way protocol for synchronization \cite{AbuSynch,SarkSynch},  prior to localization. The use of a reconfigurable intelligent surface (RIS) has been recently proposed to address some challenges in radio localization \cite{wymeersch2019radio}, along with the growing interest in communications originated by the capability of the RIS of controlling some propagation channel parameters without any need of baseband processing units \cite{ZapponeRIS,DiRenzoRIS,Guerra,Wei,Liu,Nadeem}. 

In this paper we show that, with the use of a single RIS, it is possible to perform joint localization and synchronization even with a single-antenna receiver, a single BS, and exploiting only downlink transmissions. 
A first contribution is the derivation of the direct maximum likelihood (ML) estimator for the position and clock offset; its resolution  benefits from a good initialization,  the latter typically obtained by  exhaustive  grid search.
To reduce the complexity, a second contribution  is a decoupled, relaxed estimator of position and clock offset, which does not require any a priori knowledge. The resulting  estimates  are then used to obtain a good initialization for the ML optimization. 
Results show that, even in absence of optimized beamforming and RIS control matrix, the proposed approach is able to attain the Cram\'er-Rao lower bound even for moderate values of the system parameters.

\section{System Model}

\subsection{Scenario}
\begin{figure}
    \centering
    \includegraphics[width=0.4\textwidth]{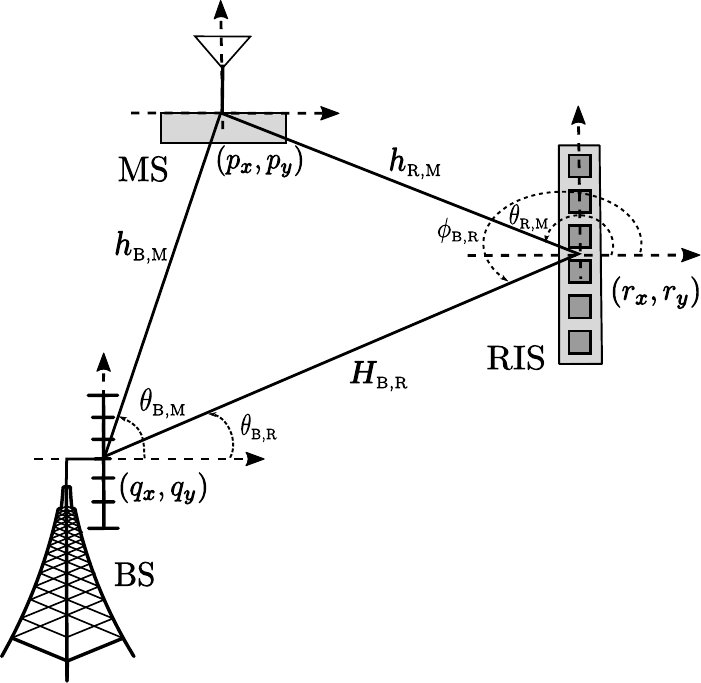}
    \caption{mmWave MISO positioning system with the aid of a RIS.}
    \label{fig:scenario}
\end{figure}

The localization scenario addressed in this paper consists of a single BS equipped with multiple antennas placed at known position $\bm{q} = [q_x \ q_y]^\mathsf{T}$, which aims at communicating with a single-antenna MS placed at unknown location $\bm{p} = [p_x \ p_y]^\mathsf{T}$, also exploiting the availability of a RIS located in the surrounding environment at a known position $\bm{r} = [r_x \ r_y]^\mathsf{T}$. More precisely, we consider a 2D scenario with uniform linear arrays (ULAs) for both the BS and the RIS elements. The numbers of antenna elements at the BS and RIS are $N_{\text{\tiny BS}}$ and $N_\R$, respectively. Fig. \ref{fig:scenario} depicts the reference scenario.

\subsection{Signal Model}
The BS transmits $G$ orthogonal frequency division multiplexing (OFDM) modulated mmWave signals over $N$ different subcarriers. Each $g$-th transmission comprises $M$ different symbols for
each subcarrier $n = 0,\ldots,N -1$, i.e., $\bm{x}^g[n] = [x^g_1[n] \cdots x^g_M[n]]^\mathsf{T} \in \mathbb{C}^{M \times 1}
$. A beamforming matrix $\bm{F} = [\bm{f}_1  \cdots  \bm{f}_M] \in \mathbb{C}^{N_{\text{\tiny BS}} \times M}$, with $\|\bm{F}\|_{\text{F}} = 1$, is applied at the BS side to precode the symbols, which are then broadcast at each transmission time $g$ over subcarrier $n$ as $\bm{s}^g[n] = \bm{F}\bm{x}^g[n]$.
The downlink received signal at the MS associated to the $g$-th transmission over $n$-th subcarrier is given by
\begin{equation}
y^g[n] = \sqrt{P} (\bm{h}^g[n])^\mathsf{T}\bm{s}^g[n] + \nu^g[n] \qquad n=0,\ldots,N-1
\end{equation}
with $P$ the transmitted power,  $\nu^g[n]$ circularly complex Gaussian noise having zero mean and variance $\sigma^2$, and $\bm{h}^g[n]$ the multipath channel.

\subsection{Channel Model}
The entire channel, including both the line-of-sight (LOS) path and the NLOS path (i.e., the reflected path via the RIS), between the BS and the MS for the $g$-th transmission and $n$-th subcarrier can be expressed as
\begin{equation}
(\bm{h}^g[n])^\mathsf{T} = \bm{h}^\mathsf{T}_{\text{\tiny{B,M}}}[n] + \bm{h}^\mathsf{T}_{\text{\tiny{R,M}}}[n]\bm{\Omega}^g\bm{H}_{\text{\tiny{B,R}}}[n]
\end{equation}
where $\bm{\Omega}^g = \mathrm{diag}(\e^{j\omega^g_1},\cdots, \e^{j\omega^g_{N_R}}) \in  \mathbb{C}^{N_R \times N_R}$ is the phase control matrix of the RIS at transmission $g$. It is a diagonal matrix with unit-modulus entries. The LOS channel $\bm{h}^\mathsf{T}_{\text{\tiny{B,M}}}[n]$ and NLOS channel components  $\bm{h}^\mathsf{T}_{\text{\tiny{R,M}}}[n]$ and $\bm{H}_{\text{\tiny{B,R}}}[n]$ are defined below. 

The direct LOS channel between the BS and MS for the $n$-th subcarrier is expressed by the complex $1\times N_\text{\tiny{BS}}$ vector
\begin{equation}
    \bm{h}^\mathsf{T}_{\text{\tiny{B,M}}}[n] = \alpha_{\text{\tiny{B,M}}}\e^{-j2\pi n \frac{\tau_{\text{\tiny{B,M}}} }{NT}}\bm{a}^\mathsf{T}_{{\text{\tiny{BS}}}}(\theta_\text{\tiny{B,M}})
\end{equation}
where $T = 1/B$ is the sampling period with $B$ being the bandwidth, $\alpha_{\text{\tiny{B,M}}} = \rho_{\text{\tiny B,M}}\e^{j\varphi_{\text{\tiny B,M}}}$ with $\rho_{\text{\tiny B,M}}$ and $\varphi_{\text{\tiny B,M}}$ denoting the modulus and phase of the complex amplitude $\alpha_{\text{\tiny{B,M}}}$, $\theta_\text{\tiny{B,M}}$ is the AOD, and $\tau_{\text{\tiny{B,M}}}$ is the 
delay including the TOF plus the clock offset $\Delta$ between the BS and the MS, as better specified later. The expression of the BS array steering vector is given by
$\bm{a}_{{\text{\tiny{BS}}}}(\theta) = [1 \ e^{j\frac{2\pi}{\lambda_c} d\sin\theta} \cdots \ e^{j(N_{\text{\tiny BS}}-1)\frac{2\pi}{\lambda_c}d\sin\theta}]^\mathsf{T}$
with $\lambda_c = c/f_c$, $f_c$ being the carrier frequency and $c$ the speed of light, and $d = \lambda_c/2$ the antenna element spacing.

The channel $\bm{H}_{\text{\tiny{B,R}}}[n] \in \mathbb{C}^{N_R \times N_\text{\tiny{BS}}}$ from  BS to  RIS is
\begin{equation}
\bm{H}_{\text{\tiny{B,R}}}[n] = \alpha_{\text{\tiny{B,R}}}\e^{-j2\pi n \frac{\tau_{\text{\tiny{B,R}}}}{NT}}\bm{a}_{{\text{\tiny{RIS}}}}(\phi_\text{\tiny{B,R}})\bm{a}^\mathsf{T}_{{\text{\tiny{BS}}}}(\theta_\text{\tiny{B,R}})
\end{equation}
where $\alpha_{\text{\tiny{B,R}}} = \rho_{\text{\tiny B,R}}\e^{j\varphi_{\text{\tiny B,R}}}$ is the complex gain over the BS-RIS path, $\phi_\text{\tiny{B,R}}$  the AOA, $\theta_\text{\tiny{B,R}}$  the AOD, and $\tau_{\text{\tiny{B,R}}}$  the TOF. The vector  $\bm{a}_{{\text{\tiny{RIS}}}}(\cdot)$ denotes the array steering vector of the RIS and it is defined in the same way as $\bm{a}_\BS(\theta)$. 
The  channel $\bm{h}_{\text{\tiny{R,M}}}[n] \in \mathbb{C}^{N_R \times 1}$ from RIS to MS is 
\begin{equation}
    \bm{h}^\mathsf{T}_{\text{\tiny{R,M}}}[n] = \alpha_{\text{\tiny{R,M}}}\e^{-j2\pi n \frac{\tau_{\text{\tiny{R,M}}} }{NT}}\bm{a}^\mathsf{T}_{{\text{\tiny{RIS}}}}(\theta_\text{\tiny{R,M}})
\end{equation}
with the notations $\alpha_{\text{\tiny{R,M}}} = \rho_{\text{\tiny R,M}}\e^{j\varphi_{\text{\tiny R,M}}}$, $\tau_{\text{\tiny{R,M}}}$, and $\theta_\text{\tiny{R,M}}$ having the same meaning as in the BS-to-MS channel model.

Finally, the geometric relationships among the BS, RIS, and MS are as follows (assuming for simplicity that the BS is placed at the origin of the reference system, i.e., $\bm{q} = [0 \ 0]^\mathsf{T}$):
\begin{align}\label{geomrelationships}
&\tau_{\text{\tiny{B,M}}} = \|\bm{p}\|/c +\Delta \nonumber \\
& \tau_{\text{\tiny{R}}} = \tau_{\text{\tiny{B,R}}} +  \tau_{\text{\tiny{R,M}}} = (\|\bm{r}\| + \|\bm{r} - \bm{p}\|)/c +\Delta \nonumber \\          &\theta_{\text{\tiny{B,M}}} = \mathrm{atan2}(p_y,p_x), \quad \theta_{\text{\tiny{R,M}}} = \mathrm{atan2}(p_y-r_y,p_x-r_x)  \nonumber \\
&\theta_{\text{\tiny{B,R}}} = \mathrm{atan2}(r_y,r_x), \quad \phi_{\text{\tiny{B,R}}} = -\pi + \theta_{\text{\tiny{B,R}}}.
\end{align}
Notice that $\tau_{\text{\tiny{B,R}}}$, $\theta_{\text{\tiny{B,R}}}$ and $\phi_{\text{\tiny{B,R}}}$ are \emph{known} quantities being the BS and RIS placed at known positions. 
The goal of this work is to solve the problem of joint localization and synchronization of the MS by exploiting the downlink signals received through both the direct LOS path and the reflected (controllable) NLOS path generated by the RIS.

\section{Cram\'er-Rao Lower Bounds}

 In this section, we briefly present the steps required to compute the fundamental bounds on the estimation of the desired MS position $\bm{p}$ and clock offset $\Delta$, to be used for the performance assessment of the proposed estimator. Due to lack of space, the explicit expressions of the involved matrices are omitted.

In the first step, we compute the FIM of the  channel parameter vector
$    \bm{\gamma} = \left[\tau_{\text{\tiny{B,M}}} \ \theta_{\text{\tiny{B,M}}} \ \rho_{\text{\tiny{B,M}}} \
    \varphi_{\text{\tiny{B,M}}} \ \tau_{\text{\tiny{R,M}}} \ \theta_{\text{\tiny{R,M}}} \ \rho_{\text{\tiny{R}}} \ \varphi_{\text{\tiny{R}}} \right]^\mathsf{T}
$
where $\rho_\text{\tiny{R}} = \rho_{\text{\tiny B,R}}\rho_{\text{\tiny R,M}}$ and $\varphi_\text{\tiny{R}} = \varphi_\text{\tiny{B,R}} + \varphi_\text{\tiny{R,M}}$. The corresponding FIM in the channel domain is given by $\displaystyle \bm{J}_{\bm{\gamma}} =  \frac{2}{\sigma^2} \sum_{g=1}^{G} \sum_{n=0}^{N-1} \Re\left\{\left(\frac{\partial m^g[n]}{\partial \bm{\gamma}}\right)^\mathsf{H} \frac{\partial m^g[n]}{\partial \bm{\gamma}}\right\}$,
with $m^g[n] = \sqrt{P} (\bm{h}^g[n])^\mathsf{T}\bm{s}^g[n]$ the noise-free version of $y^g[n]$.
We then apply a transformation of variables from the vector of the unknown channel parameters $\bm{\gamma}$ to the vector of location parameters
$
    \bm{\eta} = \left[p_x \ p_y  \ \rho_\text{\tiny B,M} \ \varphi_\text{\tiny B,M} \ \rho_\text{\tiny R} \ \varphi_\text{\tiny R} \  \Delta \right]^\mathsf{T}. 
$
The FIM of $\bm{\eta}$, denoted as $\bm{J}_{\bm{\eta}}$, is obtained by means of the transformation matrix $\bm{T}$ as
$
\bm{J}_{\bm{\eta}} = \bm{T}\bm{J}_{\bm{\gamma}}\bm{T}^\mathsf{T}     
$
where $\bm{T} \eqdef \frac{\partial \bm{\gamma}^{\mathsf{T}}}{\partial \bm{\eta}}$.

The lower bounds on the uncertainty of MS position and clock offset estimation are derived from the diagonal elements of the CRLB in the location domain, the latter obtained by inverting the corresponding FIM, i.e., $\bm{\Psi} = \bm{J}^{-1}_{\bm{\eta}}$. Specifically, the position error bound (PEB) is computed as
\begin{equation}\label{eq::PEB}
\text{PEB} = \sqrt{\left[\bm{\Psi}\right]_{1,1} + \left[\bm{\Psi}\right]_{2,2}}
\end{equation}
where $[\bm{\Psi}]_{j,j}$ selects the $j$-th diagonal entry of $\bm{\Psi}$. Similarly, the clock offset error bound (CEB) is computed as
\begin{equation}
    \text{CEB} = \sqrt{\left[\bm{\Psi}\right]_{7,7}}.
\end{equation}

\section{Maximum Likelihood Joint Localization and Synchronization}

\subsection{Joint Maximum Likelihood Estimation}
In this section, we derive the direct ML estimator of the desired  $\bm{\Theta} = [p_x \ p_y \ \Delta]^\mathsf{T}$ parameters. To this aim, we first parameterize the unknown AODs ($\theta_\BM$ and $\theta_\RM$) and TOFs ($\tau_\BM$ and $\tau_\R$) as a function of the sought $\bm{\Theta}$ through  \eqref{geomrelationships}, and then derive the optimal likelihood function in the position domain.

We stack all the $N$ samples received at each transmission $g$
    \begin{equation}\label{eq::Model1}
    \bm{y}^g = \sqrt{P}\bm{B}^g \bm{\alpha} +\bm{\nu}^g
    \end{equation}
   with 
   \begin{align*}
  \bm{y}^g &= [y^g[0] \ \cdots \ y^g[N-1]]^\mathsf{T}\\ \bm{\alpha} &=[\alpha_\BM \ \alpha_\R]^\mathsf{T}\\ \bm{B}^g &= [
(\tilde{\bm{S}}_\BM^g)^\mathsf{T}\bm{a}_\BS(\theta_\BM), \;\;  (\tilde{\bm{S}}_\R^g)^\mathsf{T}\bm{A}^\mathsf{T}(\bm{\Omega}^g)^\mathsf{T}\bm{a}_\RIS(\theta_\RM)]\\  
\tilde{\bm{S}}^g_\BM &= [\bm{s}^g[0] \;\; \cdots \;\;\e^{-j\kappa_{N-1} \tau_\text{\tiny{B,M}}}\bm{s}^g[N-1]]
\end{align*}
($\tilde{\bm{S}}_\R^g$ is obtained as $\tilde{\bm{S}}_\BM^g$ but with $\tau_\R$ in place of $\tau_\BM$),
     $\bm{A} = \bm{a}_{{\text{\tiny{RIS}}}}(\phi_\text{\tiny{B,R}})\bm{a}^\mathsf{T}_{{\text{\tiny{BS}}}}(\theta_\text{\tiny{B,R}})$, $\kappa_n = 2\pi n/NT$, and $\alpha_\R = \alpha_\BR \alpha_\RM$. By assuming that $\sigma^2$ is known (if not, it can be straightforwardly estimated as $\hat{\sigma}^2 = \sum_{g=1}^G \|\bm{y}^{g} - \sqrt{ P} \bm{B}^{g}\bm{\alpha}\|^2 / (NG)$ once the rest of parameters are obtained), $\bm{\alpha}$ remains the sole vector of unknown nuisance parameters. The ML estimation problem can be thus formulated as
    \begin{equation}
    \hat{\bm{\Theta}}^\ML = \arg \min_{\bm{\Theta}} \left[ \min_{\bm{\alpha}} L(\bm{\Theta},\bm{\alpha}) \right]  
    \end{equation}
where 
$
    L(\bm{\Theta},\bm{\alpha}) = \sum_{g=1}^G \| \bm{y}^g -\sqrt{P}\bm{B}^g \bm{\alpha} \|^2.
$
This likelihood function can be minimized with respect to $\bm{\alpha}$, yielding
$
\hat{\bm{\alpha}}^\ML = \frac{1}{\sqrt{P}} \bm{B}^{-1} \sum_{g=1}^G (\bm{B}^g)^\mathsf{H}\bm{y}^g
$
where $\bm{B} = \sum_{g=1}^G (\bm{B}^g)^\mathsf{H}\bm{B}^g$. Substituting $\hat{\bm{\alpha}}^\ML$ back into the likelihood function leads to
\begin{equation}
    L(\bm{\Theta}) = \sum_{g=1}^G \| \bm{y}^g -\sqrt{P}\bm{B}^g(\bm{\Theta}) \hat{\bm{\alpha}}(\bm{\Theta}) \|^2
\end{equation}
where we have emphasized that the only dependency left is the one on the desired parameters $\bm{\Theta}$. Accordingly, the final ML estimator is given by
\begin{equation}\label{eq::MLcost}
\hat{\bm{\Theta}}^\ML = \arg \min_{\bm{\Theta}}     L(\bm{\Theta}).
\end{equation}
Unfortunately, $L(\bm{\Theta})$ is highly non-linear and exhibits several local minima; as such, a good initialization is required to obtain $\hat{\bm{\Theta}}^\ML$ by iterative optimization. This is usually accomplished by performing an exhaustive 3D grid search over the space of the unknown $\bm{p}$ and $\Delta$. To avoid such a burden, in the next section we derive a reduced-complexity relaxed ML estimator of the position and the clock offset, able to provide a good initialization for the iterative optimization of \eqref{eq::MLcost}.

\subsection{Relaxed Maximum Likelihood  Estimation} By stacking all the observations collected over the $G$ transmissions and relaxing the dependency on the delays and AODs in (\ref{eq::Model1}), we can write 
    \begin{equation}\label{relaxedmodel}
        \underbrace{\begin{bmatrix}
        \bm{y}^1 \\
        \vdots \\
        \bm{y}^G
        \end{bmatrix}}_{\bm{y} \in \mathbb{C}^{GN\times 1}} = \underbrace{\begin{bmatrix}
        \bm{\Phi}^1_{\text{\tiny B,M}}(\theta_{\text{\tiny B,M}}(\bm{p})) & \bm{\Phi}^1_{\text{\tiny R,M}}(\theta_{\text{\tiny R,M}}(\bm{p})) \\
        \vdots & \vdots \\
        \bm{\Phi}^G_{\text{\tiny B,M}}(\theta_{\text{\tiny B,M}}(\bm{p})) & \bm{\Phi}^G_{\text{\tiny R,M}}(\theta_{\text{\tiny R,M}}(\bm{p}))
        \end{bmatrix}}_{\bm{\Phi}(\theta_\BM(\bm{p}),\theta_\RM(\bm{p})) \eqdef \bm{\Phi}(\bm{p}) \in \mathbb{C}^{GN \times 2N}}
        \underbrace{\begin{bmatrix}
        \bm{e}_{\text{\tiny B,M}} \\
        \bm{e}_{\text{\tiny R}}
        \end{bmatrix}}_{\bm{e}\in \mathbb{C}^{2N \times 1}} + \begin{bmatrix}
        \bm{\nu}^1 \\
        \vdots \\
        \bm{\nu}^G
        \end{bmatrix}
    \end{equation}
where $\bm{\Phi}^g_{\text{\tiny B,M}}(\theta_{\text{\tiny B,M}}(\bm{p})) = \mathrm{diag}(\bm{a}^\mathsf{T}_{{\text{\tiny{BS}}}}(\theta_\text{\tiny{B,M}}(\bm{p}))\bm{S}^g)$, $\bm{\Phi}^g_{\text{\tiny R,M}}(\theta_{\text{\tiny R,M}}(\bm{p})) = \mathrm{diag}(\bm{a}^\mathsf{T}_{{\text{\tiny{RIS}}}}(\theta_\text{\tiny{R,M}}(\bm{p}))\bm{\Omega}^g\bm{A}\bm{S}^g)$, $\bm{S}^g = [\bm{s}^g[0] \ \cdots \ \bm{s}^g[N-1]]$, $g=1,\ldots,G$, and
\begin{equation}\label{eq:vectore}
\bm{e}_{\text{\tiny B,M}} =  \sqrt{P}\alpha_{\text{\tiny B,M}}  \begin{bmatrix}
1 \\
\e^{-j \kappa_1 \tau_\text{\tiny{B,M}}} \\
\vdots \\
\e^{-j\kappa_{N-1}\tau_\text{\tiny{B,M}} }
\end{bmatrix}, \ \bm{e}_{\text{\tiny R}} = \sqrt{P}\alpha_{\text{\tiny R}}\begin{bmatrix}
1 \\
\e^{-j \kappa_1\tau_\text{\tiny{R}}} \\
\vdots \\
\e^{-j\kappa_{N-1}\tau_\text{\tiny{R}}}
\end{bmatrix}.
\end{equation}
Starting from this model, we first note that the only dependency of $\bm{\Phi}$ in \eqref{relaxedmodel} is on  $\bm{p}$ via the geometric relationships with the corresponding AODs. By relaxing the dependency of $\bm{e}$ on the delays $\tau_\BM$ and $\tau_\R$, and considering it as a generic unstructured $2N$-dimensional vector, a relaxed ML-based estimator (RML) of $\bm{p}$ can be derived as
\begin{equation}\label{eq::unstructML}
\hat{\bm{p}}^\RML = \arg \min_{\bm{p}} \left[ \min_{\bm{e}} \| \bm{y} - \bm{\Phi}\bm{e} \|^2\right].   
\end{equation}
The
inner minimization of \eqref{eq::unstructML} is solved by
\begin{equation}\label{eq::estvecte}
\hat{\bm{e}}^\RML = (\bm{\Phi}^\mathsf{H}(\bm{p})\bm{\Phi}(\bm{p}))^{-1} \bm{\Phi}^\mathsf{H}(\bm{p}) \bm{y}
\end{equation}
that is, the unknown vector $\bm{e}$ is estimated by pseudo-inverting the matrix $\bm{\Phi}(\bm{p})$ (the pseudo-inverse only requires $G \geq 2$ transmissions to be well-defined). Substituting this minimizing value back in \eqref{eq::unstructML} finally yields
\begin{equation}\label{eq::RML}
\hat{\bm{p}}^\RML = \arg \min_{\bm{p}} \| \bm{P}^\perp_{\bm{\Phi}(\bm{p}) } \bm{y} \|^2 
\end{equation}
where $\bm{P}^\perp_{\bm{\Phi}(\bm{p})} = \bm{I} - \bm{\Phi}(\bm{p})\left(\bm{\Phi}^\mathsf{H}(\bm{p})\bm{\Phi}(\bm{p})\right)^{-1} \bm{\Phi}^\mathsf{H}(\bm{p})$ is the orthogonal projector onto the space spanned by the columns of $\bm{\Phi}(\bm{p})$.
Remarkably, such an approach allows us to obtain an estimate of $\bm{p}$ even without any knowledge about $\Delta$, and with a complexity reduced to a 2D search instead of the 3D required by the joint ML estimator. 

As a byproduct of the above estimation of $\bm{p}$, a decoupled closed-form estimator of the unknown delays $\tau_\BM$ and $\tau_\R$ can be obtained. Specifically, $\hat{\bm{p}}^\RML$ can be plugged in \eqref{eq::estvecte} to build an estimate of $\bm{e}$. Then, the obtained $\hat{\bm{e}}^\RML$ can be divided into two subvectors, $\hat{\bm{e}}^\RML_\BM$ and $\hat{\bm{e}}^\RML_\R$, each of length $N$. By noting that the elements of both vectors (see \eqref{eq:vectore}) are discrete samples of complex exponential functions having frequencies $-\tau_\BM/(NT)$ and $-\tau_\R/(NT)$, respectively, an estimate of $\tau_\BM$ and $\tau_\R$ can be readily obtained by searching for the peaks in the FFT of the two $N$-dimensional reconstructed vectors $\hat{\bm{e}}^\RML_\BM$ and $\hat{\bm{e}}^\RML_\R$.
Based on such estimators, an  estimate of $\Delta$ can be thus obtained as 
$$
\hat{\Delta}^\RML= \frac{1}{2} \Big[\hat{\tau}^\RML_\BM - \|\hat{\bm{p}}^\RML\|/c + \hat{\tau}^\RML_\R - (\|\bm{r}\| + \| \bm{r}-\hat{\bm{p}}^\RML \|)/c \Big].
$$


\section{Simulation Results}
We consider a setup consisting of a BS placed at known position $\bm{q} = [0 \ 0]^\mathsf{T}$ m, a RIS placed at $\bm{r} = [12 \ 7]^\mathsf{T}$ m, and the MS at unknown position $\bm{p} = [5 \ 5]^\mathsf{T}$ m. As for the other parameters, we consider $f_c = 60$ GHz, $B = 40$ MHz, $G = 5$ transmissions, $N = 30$ subcarriers and $N_\BS = N_\R = 20$. The transmitted beams are $M = N_\BS/2$ and the beamforming matrix has one beam directed towards the known AOD $\theta_\BR$ and the remaining $M-1$ are set so as to provide a uniform coverage of the area. As to the phase shifts in the RIS control matrix $\bm{\Omega}^g$, they are generated as binary  random variables $w^g_i$ taking values $0$ or $\pi$ with the same probability. The channel amplitudes are generated according to the common path loss model in free space, i.e., $\rho_\BR = \lambda_c/(4\pi\|\bm{r}\|)$,   $\rho_\BM = \lambda_c/(4\pi\|\bm{p}\|)$, and $\rho_\RM = \lambda_c/(4\pi\|\bm{p}-\bm{r}\|)$, respectively. We set the clock offset to $\Delta =\frac{1}{8} \cdot NT_s \approx 100$ ns, while the transmitted power $P$ is varied in order to obtain
different ranges of the received SNR over the LOS path, defined as $\text{SNR} = 10 \log_{10} (P\rho^2_\BM/(N_0B))$, where $N_0$ is the noise power spectral density.

\begin{figure}
\centering
 \includegraphics[width=0.42\textwidth]{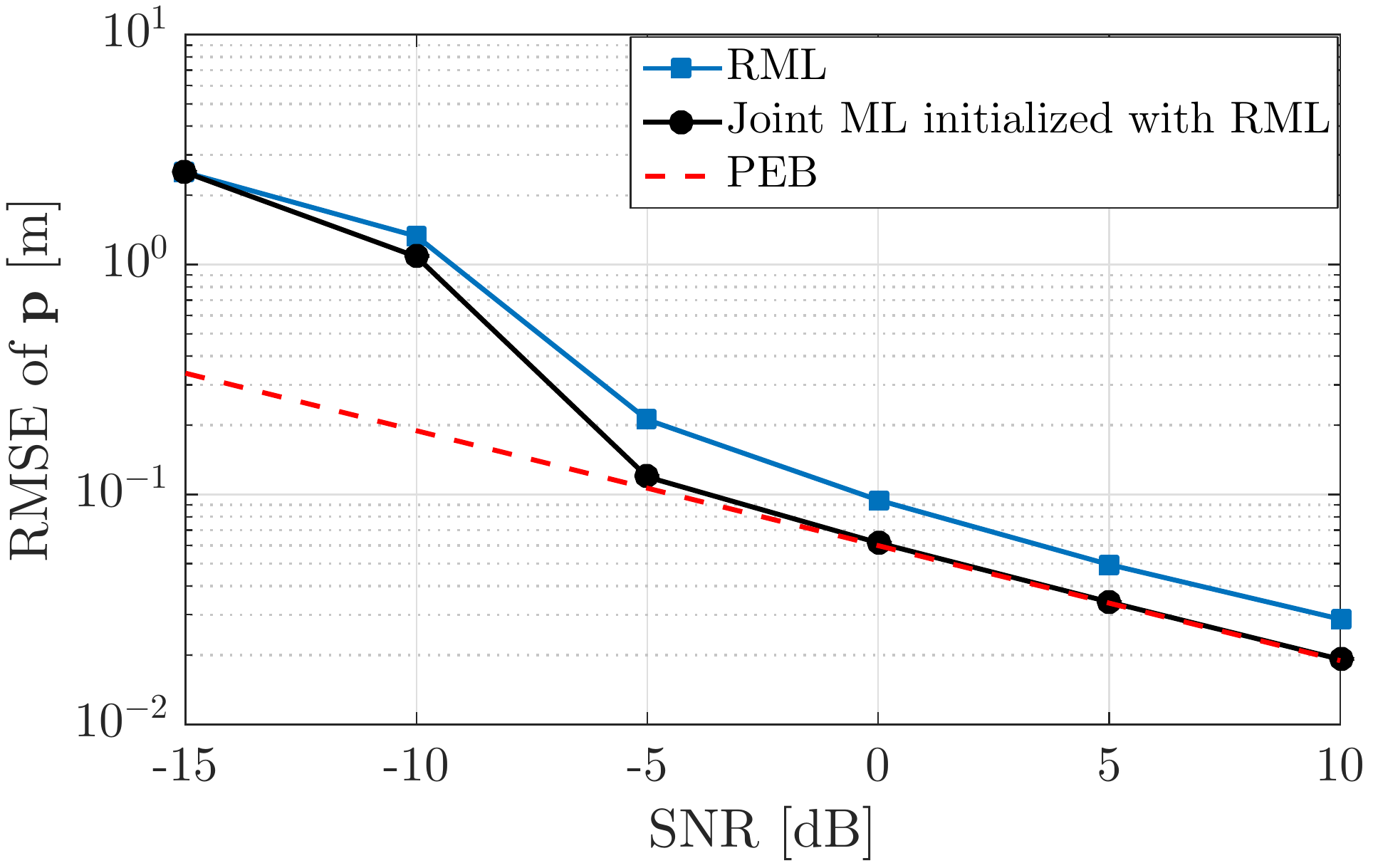}
 	\caption{RMSE on the estimation of $\bm{p}$ as a function of SNR, for the joint ML initialized with the estimates from the RML.}
\label{fig:RMSE_pos}
 \end{figure}

In Fig.~\ref{fig:RMSE_pos}, we report the root mean squared error (RMSE) on the estimation of the MS position $\bm{p}$, computed on 200 independent Monte Carlo trials, as a function of the SNR.
The proposed RML algorithm is able to achieve good estimation performance already at SNR of $-5$ dB; the RMSE then tends to decrease as the SNR further increases, but due to the intrisic suboptimality of the RML, it does not strictly attain the PEB. Conversely, the RMSE of the joint ML estimator immediately drops as soon as the initialization provided by the RML becomes accurate, achieving excellent localization performance starting from SNRs of about $-5$ dB.
 
\begin{figure}
\centering
 \includegraphics[width=0.42\textwidth]{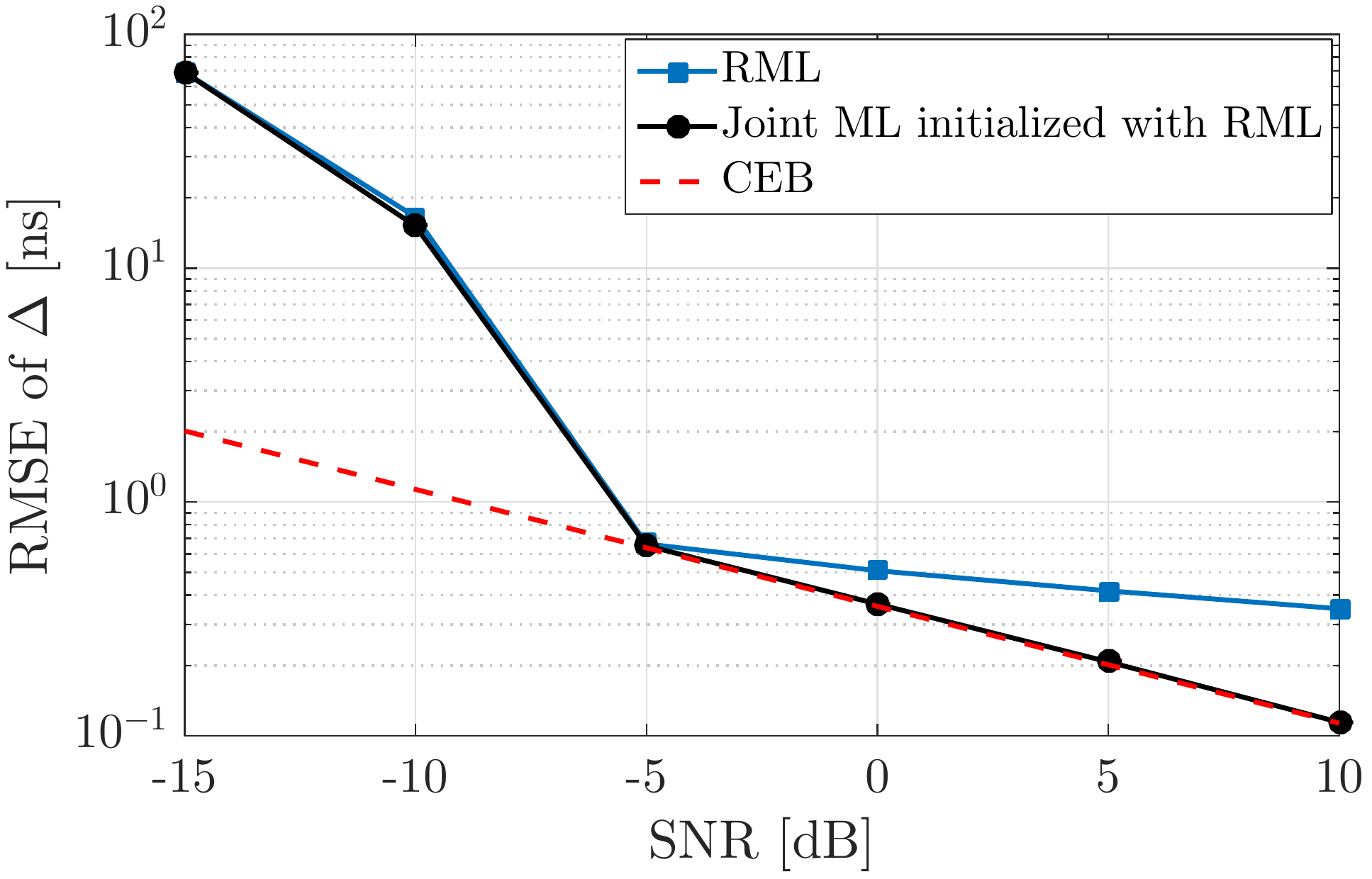}
 	\caption{RMSE on the estimation of $\Delta$ as a function of SNR, for the joint ML initialized with the estimates from the RML.}
\label{fig:RMSE_Delta}
\end{figure}

In Fig.~\ref{fig:RMSE_Delta}, we evaluate the RMSE on the estimation of the clock offset $\Delta$. 
\begin{figure}
\centering
 \includegraphics[width=0.43\textwidth]{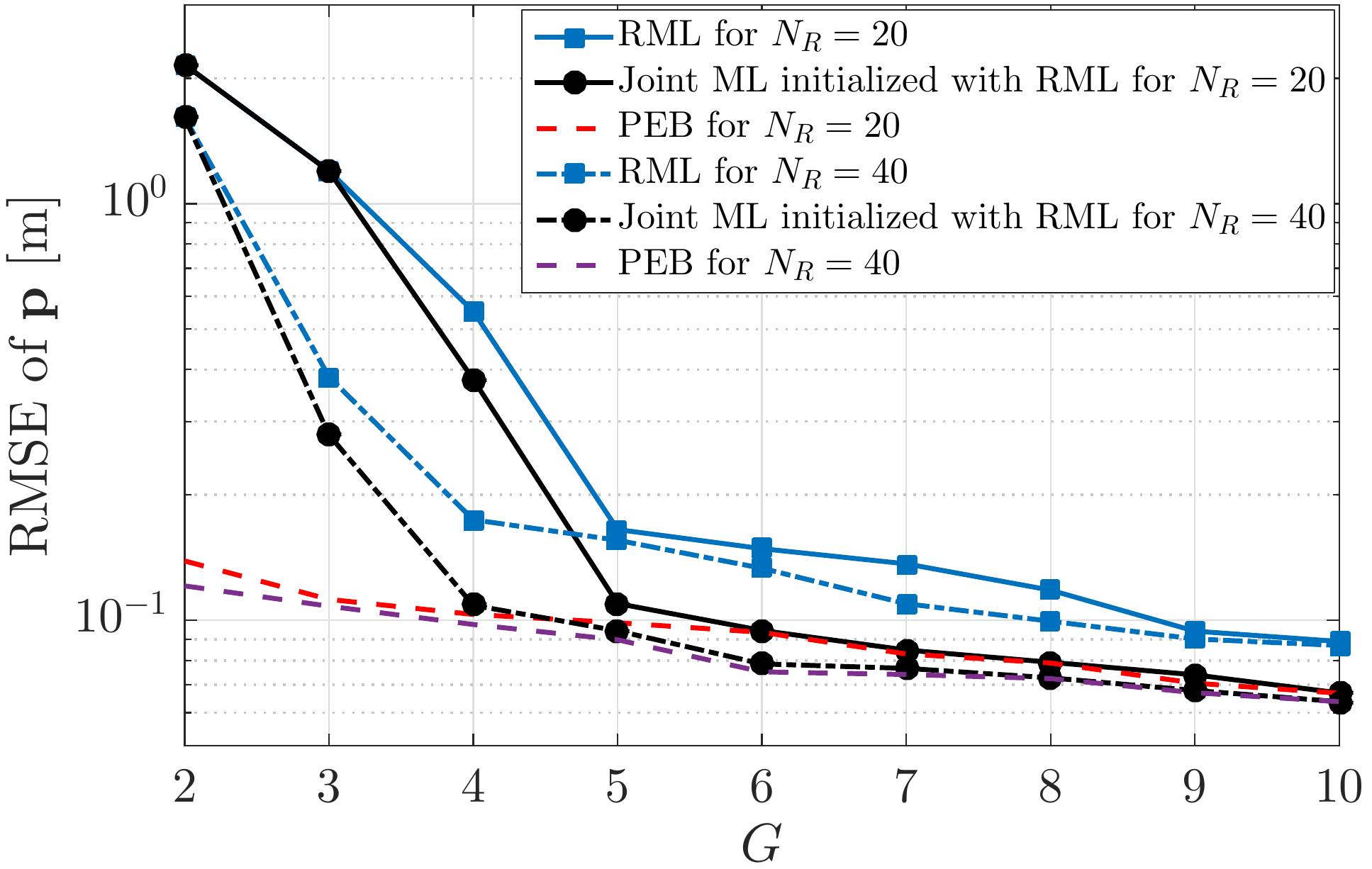}
 	\caption{RMSE on the estimation of $\bm{p}$ as a function of  $G$ for two different number of RIS elements $N_R$, at $\text{SNR}=-5$ dB.}
\label{fig:RMSE_Ganalysis}
\end{figure}
Focusing on the RMSE of the RML, we notice a similar decreasing trend as the one experienced in the low SNR regime of Fig.~\ref{fig:RMSE_pos}. In this case, however, the gap with the CEB in the moderate to high SNR regime is slightly wider, given that the estimate of $\Delta$ is based on the suboptimal RML followed by the application of another suboptimal (though closed-form) estimation approach such as the FFT. Remarkably, the joint ML estimator is able to attain the bound for even low values of the SNR.

To further corroborate the above analysis, we investigate the impact of the number of transmissions $G$ on the PEB for the challenging case of $\text{SNR} = -5$ dB, comparing the scenarios with $N_\R=20$ and $N_\R=40$ elements on the RIS. By looking at the solid curves reported in Fig.~\ref{fig:RMSE_Ganalysis} it emerges that, although the pseudo-inverse required to implement the RML estimator in \eqref{eq::RML} is well defined already for $G \geq 2$, at least $G = 5$ transmissions are needed by the joint ML to achieve the PEB. Interestingly, when the number of RIS elements is doubled ($N_\R =40$), the joint ML estimator is able to reach high accuracy already with $G = 4$ transmissions, in spite of the low SNR experienced by the received signals. Another interesting insight follows from a direct comparison between the PEBs in the two considered cases: indeed, it is evident that the theoretical error is only slightly reduced as $N_\R$ increases from 20 to 40. This behavior is linked to the fact that both the transmit beamforming as well as the RIS control matrix are not optimized for the localization task (a wider gap between the PEBs can be expected in case of optimized parameters); thus, in the case of random RIS weights, a reduction of the number of RIS elements can be compensated by increasing the number of transmissions.


\section{Conclusion}
We have addressed the problem of joint localization and synchronization of a MS in a mmWave MISO system with the aid of a RIS.
First, the joint ML estimation problem has been formulated in the position domain, and its solution formally derived. To overcome the need of an exhaustive three-dimensional search to initialize the iterative optimization of the joint ML estimator, a reduced-complexity decoupled estimator of the position and clock offset based on a proper relaxation of the signal model has been proposed. Simulation results have demonstrated that the proposed approach can achieve high localization and synchronization accuracy already for moderate values of the system parameters, even in the low SNR regime, without optimizing neither the transmit beamforming nor the RIS control matrix, and without any a priori knowledge of the clock offset. 



\bibliographystyle{IEEEbib}
\bibliography{refs}

\end{document}